\documentclass[%
superscriptaddress,
nofootinbib,
nobibnotes,
amsmath,amssymb,
aps,
floatfix]{revtex4-2}

\usepackage{placeins}
\usepackage{booktabs}
\usepackage{dcolumn}
\usepackage{array}
\usepackage{longtable}
\usepackage{float}
\usepackage{hyphenat}
\usepackage{natbib}
\usepackage{hyperref}

\usepackage{amsmath,amsfonts,amsthm,amssymb}
\usepackage{bm,graphicx,graphics,color}
\usepackage{epsf,epsfig}
\usepackage[all]{xy}
\newcommand*{\B}[1]{\ifmmode\bm{#1}\else\textbf{#1}\fi}

\def\bx{\mathbf{x}}

\def\bv{\mathbf{v}}

\def\bk{\mathbf{k}}

\def\no{\nonumber}
\def\lb{\label}
\def\be{\begin{equation}}
\def\ee#1{\label{#1}\end{equation}}
\newcommand{\ben}{\begin{eqnarray}}
\newcommand{\een}{\end{eqnarray}}
\usepackage[english]{babel}

\usepackage[letterpaper,top=2cm,bottom=2cm,left=3cm,right=3cm,marginparwidth=1.75cm]{geometry}

\usepackage{amsmath}
\usepackage{graphicx}
\begin{document}

\title{A self-gravitating system composed of baryonic and  dark matter analysed from the post-Newtonian Boltzmann equations}

\author{Gilberto M. Kremer}
\email{kremer@fisica.ufpr.br}
\affiliation{Departamento de F\'{i}sica, Universidade Federal do Paran\'{a}, Curitiba 81531-980, Brazil}
\author{Kamel Ourabah}
\email{kam.ourabah@gmail.com}
\affiliation{Theoretical Physics Laboratory, Faculty of Physics, University of Bab-Ezzouar, USTHB, Boite Postale 32, El Alia, Algiers 16111, Algeria}

\begin{abstract}
We study the Jeans gravitational instability for a mixture of baryonic and dark matter particles, in the post-Newtonian approximation. We adopt a kinetic model consisting of a coupled system of post-Newtonian collisionless Boltzmann
equations, for each species, coupled to the post-Newtonian Poisson equations. We derive the stability criterion, accounting for both post-Newtonian corrections and the presence of dark matter. It is shown that both effects give rise to smaller Jeans masses, in comparison with the standard Jeans criterion, meaning that a smaller mass is needed to begin the gravitational collapse. Taking advantage of that, we confront the model with the observational stability of Bok globules, and show that the model correctly reproduces the data. 
\end{abstract}

\maketitle

\section{Introduction}

The analysis of instabilities of self-gravitating fluids is an old subject in the literature which goes back to the pioneer work of Jeans  \cite{Jeans}, who determined from the hydrodynamic equations coupled with the Newtonian Poisson equation a dispersion relation where one solution is interpreted as a growth of  mass density perturbations in time.

One is referred to the books \cite{Wein,Coles,BT1} for a description of the  mass density perturbations which grow exponentially with time -- known as Jeans instability -- which is associated with the  gravitational collapse of  self-gravitating interstellar gas clouds,
where the outwards pressure force becomes smaller than the inwards gravitational force.

Although formulated 120 years ago, the process of Jeans instability still constitutes an active area of research, revisited from various modern standpoints. 
One may cite for instance its generalisation to general relativity \cite{a2} and to an expanding universe background \cite{a2,a3,a4}, its formulation in the language of kinetic theory \cite{a5,a6}, and its generalisation to alternative theories of gravity \cite{a7,a8,a9,a10}, where it is regularly used to constraint the free parameters of the theory. 

Presently the matter content of the Universe is known to be composed by baryonic matter -- which consists of all categories of atoms -- and dark matter -- a still unknown component which does not interact with electromagnetic radiation. Cold dark matter is important in the structure formation, since it interacts only with gravity, collapses earlier and forms the seeds where baryons fall later. The formation of structures would occur later if the cold dark matter was not present.

The Jeans instability for a system composed by baryonic and dark matter particles was investigated within the framework of a coupled system of collisionless Boltzmann equations and the Newtonian Poisson equation in \cite{b5,b6,b8} and by using a hybrid quantum-classical fluid approach in \cite{our}.

The Jeans instability  was also investigated on the basis  of the first post-Newtonian  hydrodynamic equations \cite{NKR,NH} and of the first  \cite{K1} and second \cite {K2} post-Newtonian Boltzmann equations  coupled with the Newtonian and the post-Newtonian Poisson equations.

In the present work  the Jeans instability for a system consisted of baryonic and cold dark matter is analysed within the framework of   the system of first post-Newtonian collisionless Boltzmann equations which are coupled with  the Newtonian and  post-Newtonian Poisson equations.  From a perturbation analysis of the one-particle distribution functions and gravitational potentials in terms of plane wave representations a dispersion relation is obtained and the post-Newtonian Jeans mass is derived. As an application of the post-Newtonian Jeans mass for the system of baryons and dark matter the observational stability data of Bok globules   is investigated.

 The paper has the following structure: in Section \ref{sec2} the coupled system of post-Newtonian Boltzmann and  Poisson equations  are introduced  as well as the equilibrium Maxwell-J\"uttner distribution functions.  The perturbations from equilibrium background states of the one-particle distribution functions and of the gravitational potentials are the subject of Section \ref{sec3}. In Section \ref{sec4} the perturbations are represented as plane waves of small amplitudes  from  which a dispersion relation  is derived and the influence of the post-Newtonian approximation in the Jeans mass of the baryonic and dark matter system    is investigated . The application of the theoretical prediction for the post-Newtonian Jeans mass  with the observational stability data of Bok globules is the topic of Section \ref{sec5}. In Section \ref{sec6} the conclusions of the work are stated.

\section{The system of Boltzmann equations}\lb{sec2}

We are interested in analysing a system composed of baryonic and  dark matter particles within the framework of collisionless post-Newtonian Boltzmann equations, since   dark matter interacts with other particles only through gravity. For that end we introduce the subscripts $b$  and $d$ to denote the baryonic and dark matter, respectively, so that the post-Newtonian Boltzmann equation for the one-particle distribution function $f_\alpha=f(\bx,\bv_\alpha,t)$ of the constituent  $\alpha=b,d$ -- defined in the phase space spanned by the spatial coordinates $\bx$ and particle three-velocity $\bv_\alpha$ -- reads \cite{Ped,PGMK,GGKK,K1} 
\ben\no
&&\frac{\partial f_\alpha}{\partial t}+v^\alpha_i\frac{\partial f_\alpha}{\partial x^i}+\frac{\partial U}{\partial x^i}\frac{\partial f_\alpha}{\partial v^\alpha_i}
+\frac1{c^2}\bigg[\left(v_\alpha^2-4U\right)\frac{\partial U}{\partial x^i}-4v^\alpha_iv^\alpha_j\frac{\partial U}{\partial x^j}-3v^\alpha_i\frac{\partial U}{\partial t}
+2\frac{\partial \Phi}{\partial x^i}+\frac{\partial \Pi_i}{\partial t}
\\\lb{m1}
&&\qquad+v^\alpha_j\left(\frac{\partial \Pi_i}{\partial x^j}-\frac{\partial \Pi_j}{\partial x^i}\right)\bigg]\frac{\partial f_\alpha}{\partial v^\alpha_i}=0.
\een
Here the Newtonian $U$ and the post-Newtonian $\Phi$ and $\Pi_i$ gravitational potentials satisfy the Poisson equations
\ben\lb{m2a}
&&\nabla^2U=-\frac{4\pi G}{c^2} \left({\buildrel\!\!\!\! _{0} \over{T_b^{00}}}+{\buildrel\!\!\!\! _{0} \over{T_d^{00}}}\right),
\\\lb{m2b}
&& \nabla^2\Phi=-2\pi G \left({\buildrel\!\!\!\! _{2} \over{T_b^{00}}}+{\buildrel\!\!\!\! _{2} \over{T_b^{ii}}}+{\buildrel\!\!\!\! _{2} \over{T_d^{00}}}+{\buildrel\!\!\!\! _{2} \over{T_d^{ii}}}\right),
\\\lb{m2c}
&&\nabla^2\Pi^i=-\frac{16\pi G}{c}\left({\buildrel\!\!\!\! _{1} \over{T_b^{0i}}}+{\buildrel\!\!\!\! _{1} \over{T_d^{0i}}}\right)+\frac{\partial^2U}{\partial t\partial x^i}.
\een
Above $G$ is the universal gravitational constant and we have denoted the $1/c^n$-order  of the energy-momentum tensor of constituent $\alpha$ by ${\buildrel\!\!\!\! _{n} \over{T_\alpha^{\mu\nu}}}$.

The energy-momentum tensor of constituent $\alpha$ is defined in terms of the one-particle distribution function $f_\alpha$ and of the particle four-velocity  $u_\alpha^\mu$ by \cite{CK,GGKK}
\ben\lb{m3a}
T_\alpha^{\mu\nu}=m_\alpha^4c\int u_\alpha^\mu u_\alpha^\nu f_\alpha\frac{\sqrt{-g}\,d^3 u_\alpha}{u^\alpha_0},
\een
where  $m_\alpha$ denotes the rest mass of a particle of constituent $\alpha$.

The components of the $\alpha$ constituent particle four-flow in the first post-Newtonian approximation read \cite{Ch1,GGKK}
\ben\lb{m3b}
u_\alpha^0=c\left[1+\frac1{c^2}\left(\frac{v_\alpha^2}2+U\right)\right],\qquad u_\alpha^i=\frac{u_\alpha^0v_\alpha^i}c.
\een

For a relativistic gas the one-particle distribution function at equilibrium is determined by the Maxwell-J\"uttner distribution function (see e.g. \cite{CK}).  In a stationary equilibrium  background where the hydrodynamic velocity vanishes the Maxwell-J\"uttner distribution function for the constituent $\alpha$  in the first post-Newtonian approximation  is given by \cite{KRW}
\ben\lb{m4a}
f^\alpha_{MJ}=f^\alpha_0
\left\{1-\frac{\sigma_\alpha^2}{c^2}\left[\frac{15}8+\frac{3v_\alpha^4}{8\sigma_\alpha^4}
+\frac{2U v_\alpha^2}{\sigma_\alpha^4}\right]\right\},
\een
where $f^\alpha_0$ is the Maxwellian distribution function 
\ben\lb{m4b}
f^\alpha_0=\frac{\rho^\alpha_0}{m_\alpha^4(2\pi \sigma_\alpha^2)^\frac32}e^{-\frac{v_\alpha^2}{2\sigma_\alpha^2}},
\een
which is given in terms of the mass density $\rho_0^\alpha$, the gas particle three-velocity $\bf v_\alpha$, and the dispersion velocity  $\sigma_\alpha$ of constituent $\alpha$. 

The expression for the invariant integration element of the  energy-momentum tensor (\ref{m3a}) in the  first post-Newtonian approximation reads  \cite{KRW}
\ben\lb{m4c}
\frac{\sqrt{-g}\, d^3 u_\alpha}{u_0^\alpha}=
\left\{1+\frac1{c^2}\left[2v_\alpha^2+6U\right]\right\}\frac{d^3v_\alpha}c.
\een

\section{Field perturbations}\label{sec3}

In this section we shall consider perturbations from equilibrium background states of the one-particle distribution functions and of the gravitational potentials. The subscripts zero will denote the background states and the subscripts one the perturbed states, namely
\ben\lb{m6a}
&&f(\bx,\bv_\alpha,t)=f^\alpha_{MJ}(\bx,\bv_\alpha,t)+ f_1^\alpha(\bx,\bv_\alpha,t),\quad\alpha=b,d
\\ \lb{m6b}
&&U(\bx,\bv,t)=U_0(\bx)+ U_1(\bx,\bv,t),
\\\lb{m6c}
&&\Phi(\bx,\bv,t)=\Phi_0(\bx)+ \Phi_1(\bx,\bv,t),
\\\lb{m6d}
&&\Pi_i(\bx,\bv,t)=\Pi_i^0(\bx)+\Pi_i^1(\bx,\bv,t).
\een

We begin by introducing the representations    (\ref{m6a}) --  (\ref{m6d}) into the Boltzmann equation for the constituent $\alpha$ (\ref{m1}) and  get that the resulting background equation is identically satisfied if $\nabla U_0=0$, $\nabla\Phi_0=0$ and $\nabla \Pi_i^0=0$, while the perturbed equation reduces to
\ben\no
\frac{\partial f_1^\alpha}{\partial t}+v^\alpha_i\frac{\partial f^\alpha_1}{\partial x^i}+\frac{\partial U_1}{\partial x^i}\frac{\partial f_{MJ}^{0\alpha}}{\partial v_\alpha^i}-\frac{2v_\alpha^2f^\alpha_0}{\sigma_\alpha^2c^2}\bigg(\frac{\partial U_1}{\partial t}+v^\alpha_i\frac{\partial U_1}{\partial x^i}\bigg)+\frac1{c^2}\Bigg[\left(v_\alpha^2-4U_0\right)\frac{\partial U_1}{\partial x^i}+2\frac{\partial \Phi_1}{\partial x^i}
\\\lb{m7a}
-3v^\alpha_i\frac{\partial U_1}{\partial t}+\frac{\partial \Pi^1_i}{\partial t}-4v^\alpha_iv^\alpha_j\frac{\partial U_1}{\partial x^j}+v^\alpha_j\left(\frac{\partial \Pi^1_i}{\partial x^j}-\frac{\partial \Pi^1_j}{\partial x^i}\right)\Bigg]\frac{\partial f^\alpha_0}{\partial v_\alpha^i}=0.\quad
\een
Here $f_{MJ}^{0\alpha}$ is the background Maxwell-J\"uttner distribution function 
\ben\lb{m7b}
f_{MJ}^{0\alpha}=f_0^\alpha
\left\{1-\frac{\sigma_\alpha^2}{c^2}\left[\frac{15}8+\frac{3v_\alpha^4}{8\sigma_\alpha^4}
+2\frac{U_0 v_\alpha^2}{\sigma_\alpha^4}\right]\right\}.
\een

For the Poisson equations  (\ref{m2a}) and (\ref{m2b})  we assume  "Jeans  swindle" (see e.g. \cite{Coles,BT1})     and consider that they are valid only for the perturbed gravitational potentials and distribution functions. First we have to evaluate the components of the energy-momentum tensor of each constituent and next to insert the perturbed values into the Poisson equations which lead to
\ben\lb{m8a}
&&\nabla^2U_1=-4\pi G \sum_{\alpha=b}^dm_\alpha^4\int f^\alpha_1 d^3v_\alpha,
\\\lb{m8b}
&&\nabla^2\Pi_1^i=-16\pi G \sum_{\alpha=b}^dm_\alpha^4\int v^\alpha_if^\alpha_1d^3v_\alpha
+\frac{\partial^2U_1}{\partial t\partial x^i},
\\\lb{m8c}
&& \nabla^2\Phi_1=
- 2\pi G \sum_{\alpha=b}^dm_\alpha^4\int\big(4v_\alpha^2+8U_0\big)f^\alpha_1d^3v_\alpha
+4\pi G\sum_{\alpha=b}^d\rho_0^\alpha U_1.
\een

\section{Plane wave representations}\lb{sec4}

We represent the instabilities  as plane waves of frequency $\omega$,  wave number vector $\bk$ and small amplitudes $\overline f^\alpha_1, \overline U_1,\overline\Phi_1$ and $\overline{\Pi_1^i}$:
\ben\lb{m9a}
f^\alpha_1(\bx,\bv,t)=\overline f^\alpha_1e^{i(\bk\cdot\bx-\omega t)},\quad U_1(\bx,\bv,t)=\overline U_1e^{i(\bk\cdot\bx-\omega t)},
\\\lb{m9b}
\Phi_1(\bx,\bv,t)=\overline \Phi_1e^{i(\bk\cdot\bx-\omega t)},\quad \Pi^i_1(\bx,\bv,t)=\overline {\Pi_i^1}e^{i(\bk\cdot\bx-\omega t)}.
\een

From the insertion of the above plane wave representations into  the perturbed Boltzmann equation for constituent $\alpha$ (\ref{m7a}) we get the following equation which gives the perturbed amplitude of the  distribution function of constituent $\alpha$ in terms of the perturbed gravitational potentials
\ben\no
({\bf v_\alpha\cdot k}-\omega)\overline f^\alpha_1-\frac{f_0^\alpha}{\sigma_\alpha^2}\bigg\{({\bf v_\alpha\cdot k})\overline U_1\bigg[1-\frac{\sigma_\alpha^2}{c^2}\bigg(\frac{15}8+\frac{3v_\alpha^4}{8\sigma_\alpha^4}-\frac{v_\alpha^2}{2\sigma_\alpha^2}+\frac{2v_\alpha^2U_0}{\sigma_\alpha^4}\bigg)\bigg]
\\\lb{m10}
+\frac1{c^2}\bigg[v_\alpha^2\omega\overline U_1+2({\bf v_\alpha\cdot k})\overline\Phi_1
-\omega v^\alpha_i\overline{\Pi^1_i}\bigg]\bigg\}=0.
\een

Without loss of generality we consider  the wave number vector in the $x$-direction, i.e.,  ${\bf k}=(\kappa,0,0)$ and insert the perturbed amplitude of the  distribution function of constituent $\alpha$ from (\ref{m10}) and the plane wave representations of the gravitational potentials into the  Poisson  equations (\ref{m8a}) -- (\ref{m8c})  and integrate the resulting equations, yielding
\ben\no
&&\kappa^2\overline{\Pi_x^1}=\frac{16\pi G\omega}{\kappa}\sum_{\alpha=b}^d\frac{\rho^\alpha_0}{\sigma_\alpha^2}\bigg\{\bigg[I^\alpha_2-\frac{3\sigma_\alpha^2}{2c^2}\left(I^\alpha_6+\frac54I^\alpha_2\right)-\frac{4U_0}{c^2}\left(I^\alpha_2+I^\alpha_4\right)\bigg]\overline U_1
\\\lb{m11a}
&&\qquad+\frac{I^\alpha_2}{c^2}\left[2\overline\Phi_1-\frac{\omega}{\kappa}\overline{\Pi_x^1}\bigg]\right\}-\kappa\omega\overline U_1,
\\\no
&&\kappa^2\overline U_1 =4\pi G\sum_{\alpha=b}^d\frac{\rho^\alpha_0}{\sigma_\alpha^2}\bigg\{\bigg[I^\alpha_2+\left(I^\alpha_0+I^\alpha_2\right)\frac{\omega^2}{c^2\kappa^2}
-\frac{3\sigma_\alpha^2}{2c^2}\bigg(I^\alpha_6+\frac43I^\alpha_4+\frac{31}{12}I^\alpha_2\bigg)-\frac{4U_0}{c^2}\left(I^\alpha_2+I^\alpha_4\right)\bigg]\overline U_1
\\\lb{m11b}
&&\qquad+\frac{I^\alpha_2}{c^2}\bigg[2\overline\Phi_1-\frac{\omega}{\kappa}\overline{\Pi_x^1}\bigg]\bigg\},
\\\no
&&\kappa^2\overline\Phi_1 =4\pi G\sum_{\alpha=b}^d\rho^\alpha_0\overline U_1+16\pi G\sum_{\alpha=b}^d\rho^\alpha_0\bigg\{2\bigg(I^\alpha_0+I^\alpha_2
+\frac{I^\alpha_4}2\bigg)\frac{\omega^2}{\kappa^2c^2}+I^\alpha_2+I^\alpha_4-\frac{3\sigma_\alpha^2}{2c^2}\bigg(I^\alpha_8+\frac73I^\alpha_6
\\\no
&&\qquad+\frac{71}{12}I^\alpha_4+\frac{71}{12}I^\alpha_2\bigg)+\frac{U_0}{\sigma_\alpha^2}\bigg[I^\alpha_2-\frac{\sigma_\alpha^2}{c^2}\bigg(\frac{11}2I^\alpha_6+10I^\alpha_4+\frac{95}8I^\alpha_2\bigg)-\frac{4U_0}{c^2}(I^\alpha_2+I^\alpha_4)
\\\lb{m11c}
&&\qquad+\frac{\omega^2}{c^2\kappa^2}(I^\alpha_0+I^\alpha_2)\bigg]\bigg\}\overline U_1+\frac{16\pi G}{c^2}\sum_{\alpha=b}^d\rho_0^\alpha\left(I^\alpha_2+I^\alpha_4+I^\alpha_2\frac{U_0}{\sigma_\alpha^2}\right)\left[2\overline\Phi_1-\frac{\omega}{\kappa}\overline{\Pi_x^1}\right].
\een
The components $\overline{\Pi^1_y}$ and $\overline{\Pi^1_z}$ vanish and above we have introduced the integrals
\ben\lb{m11d}
I_n^\alpha(\kappa,\omega)=\frac2{\sqrt\pi}\int_0^\infty\frac{x^ne^{-x^2}}{x^2-(\omega/\sqrt2\sigma_\alpha \kappa)^2}dx, 
\een
where $x=v_x^\alpha/\sqrt2\sigma_\alpha$.

Equations  (\ref{m11a}) --  (\ref{m11c}) for the  amplitudes $\overline{\Pi_x^1}$, $\overline U_1$ and $\overline\Phi_1$ compose  an algebraic system of equations which  admits a solution  if the determinant of the coefficients of the amplitudes vanishes.  Up to the $1/c^2$-order it follows the dispersion relation
\ben\no
\kappa_*^4-\kappa_*^2\bigg\{I_2^d\bigg(1+\frac{I_2^b\rho_0^b\sigma_d^2}{I_2^d\rho_0^d\sigma_b^2}\bigg)+\frac{\sigma_d^2}{c^2}\bigg[I_2^d\bigg(1+\frac{I_2^b\rho_0^b\sigma_d^2}{I_2^d\rho_0^d\sigma_b^2}\bigg)\bigg(\frac{33}8
+4\frac{U_0}{\sigma_d^2}\bigg)+6I_4^d\bigg(1+\frac{I_4^b\rho_0^b}{I_4^d\rho_0^d}\bigg)
\\\no
-\frac32I_6^d\bigg(1+\frac{I_6^b\rho_0^b}{I_6^d\rho_0^d}\bigg)-4\frac{U_0}{\sigma_d^2}I_4^d\bigg(1+\frac{I_4^b\rho_0^b\sigma_d^2}{I_4^d\rho_0^d\sigma_b^2}\bigg)\bigg]\bigg\}-\frac{\sigma_d^2}{c^2}\bigg[I_0^d\omega^2_*\bigg(1+\frac{I_0^b\rho_0^b\sigma_d^2}{I_0^d\rho_0^d\sigma_b^2}\bigg)
\\\lb{m18}
+2I_2^d\bigg(1+\frac{\rho_0^b}{\rho_0^d}-\omega^2_*\bigg)\bigg(1+\frac{I_2^b\rho_0^b\sigma_d^2}{I_2^d\rho_0^d\sigma_b^2}\bigg)\bigg]=0.\qquad
\een
The dispersion relation relates the the dimensionless frequency $\omega_*=\omega/{\sqrt{4\pi G\rho^d_0}}$ with the dimensionless wavenumber $\kappa_*={\kappa}/{\kappa_J}$, which are given in terms of the dark matter Jeans wave number  defined by $\kappa_J=\sqrt{4\pi G\rho^d_0}/\sigma_d$.
The reason to take the dark matter to build the dimensionless quantities is that the dark matter begins to collapse into a complex network of dark matter halos well before the ordinary matter.

From the analysis of the dispersion relation algebraic equation (\ref{m18}) we infer two distinct regimes: in one of them the frequency assumes real values implying that the perturbations will propagate as harmonic waves in time, while in the other the frequency has pure imaginary values and the perturbations will grow or decay in time. 

Here we are interesting in analysing the case where  an instability happens  which corresponds to the Jeans instability and is related to the  minimum mass where an overdensity begins the gravitational collapse.  In this case the limiting value of the frequency where  the instability occurs is when  $\omega_*=0$ and the dispersion relation (\ref{m18}) reduces to
\be
\kappa_*^4-\bigg(1+\frac{\rho_0^b\sigma_d^2}{\rho_0^d\sigma_b^2}\bigg)\kappa_*^2+2\frac{\sigma_d^2}{c^2}\bigg\{1+3\bigg(1+\frac{\rho_0^b}{\rho_0^d}\bigg)\kappa_*^2+\frac{\rho_0^b}{\rho_0^d}\bigg[1
+\frac{\sigma_d^2}{\sigma_b^2}\bigg(1+\frac{\rho_0^b}{\rho_0^d}\bigg)\bigg]+\frac{U_0}{\sigma_d^2}\bigg(1+\frac{\rho_0^b\sigma_d^2}{\rho_0^d\sigma_b^2}\bigg)\kappa_*^2\bigg\}
=0,
\ee{m20a}
since  the integrals have the values $I_2^b=I_2^d=1$, $I_4^b=I_4^d=1/2$ and  $I_6^b=I_6^d=3/4$.

The  minimum mass for an overdensity to star the gravitational collapse is related with the real positive value of $\kappa_*$ obtained from (\ref{m20a}), which by considering  terms up to the $1/c^2$ order  reads
\ben\lb{m20b}
\kappa_*=\bigg(1+\frac{\rho_0^b\sigma_d^2}{\rho_0^d\sigma_b^2}\bigg)^{-\frac12}\bigg\{1+\frac{\sigma_d^2}{c^2}\bigg(4+\frac{U_0}{\sigma_d^2}\bigg)
+\frac{\rho_0^b\sigma_d^2}{\rho_0^d\sigma_b^2}\bigg[1+\frac{\sigma_b^2}{c^2}\bigg(4+\frac{U_0}{\sigma_b^2}\bigg)\bigg]\bigg\}.
\een

From (\ref{m20b}) two limiting cases are interesting to analyse. The first one is when we have only one component present, which we choose as the dark matter component, in this case we take a vanishing  mass density of the baryonic matter $\rho_0^b=0$ and get
\ben\lb{m20c}
\kappa_*=1+\frac{\sigma_d^2}{c^2}\bigg(4+\frac{U_0}{\sigma_d^2}\bigg),
\een
which is the expression given in \cite{K1}.

On the other hand without the relativistic correction we have from (\ref{m20b}) by neglecting the $1/c^2$ terms 
\ben\lb{m20d}
\kappa_*=\sqrt{1+\frac{\rho_0^b\sigma_d^2}{\rho_0^d\sigma_b^2}},
\een
and we recover the result given in \cite{b5,b6}.

The Jeans mass is associated with the mass contained in a sphere of radius equal to the wavelength of the mass perturbation. By building the ratio of the Jeans masses corresponding to the dark-baryonic system $M_J^{db}$ and the dark matter system $M_J^d$ we get up to $1/c^2$ terms 
\ben\lb{m21}
\frac{M_J^{db}}{M_J^{d}}=\bigg(1+\frac{\rho_0^b}{\rho_0^d}\bigg)\bigg({1+\frac{\rho_0^b\sigma_d^2}{\rho_0^d\sigma_b^2}}\bigg)^{-\frac52}\bigg\{1-3\frac{\sigma_d^2}{c^2}\bigg(\frac{U_0}{\sigma_d^2}+4\bigg)
+\frac{\rho_0^b\sigma_d^2}{\rho_0^d\sigma_b^2}\bigg[1-3\frac{\sigma_d^2}{c^2}\bigg(\frac{U_0}{\sigma_d^2}+4\frac{\sigma_b^2}{\sigma_d^2}\bigg)\bigg]\bigg\}.
\een

\section{Comparison with observational data}\lb{sec5}

To assess the physical viability of the present model, it is interesting to confront it with the observation of regions in the Universe that can experience star formation. 
Here, we compare the theoretical prediction for the Jeans mass (\ref{m21}), with the observational stability data of Bok globules. The latter are nearby isolated clouds of interstellar gas and dust with simple shapes. They have characteristic temperatures of the order of $10K$, and masses of $\sim 10 M_{\odot}$, which are close to their corresponding Jeans masses. This last characteristic places them among the most interesting astrophysical objects to test a deviation from the standard Jeans criterion, since a small modification of their Jeans mass leads to a different prediction for their stability.

We are mainly interested in the data of \cite{Kan} (see also \cite{Vain}), consisting of a set of 11 Bok globules, reproduced in Table \ref{tab:globules}, together with their kinetic temperature, density, mass, Jeans mass (assuming Newtonian gravity and in the absence of a dark matter background), and their observed stability.

\begin{table}[h!]
\scriptsize
\begin{tabular}{|l|c|c|c|c|c|c|c|} \hline
Bok Globule &$T\text{[K]}$ & $n_{H_2}\text{[cm$^{-3}$]}$ & $M [M_{\odot}]$  & $M_J [M_{\odot}]$ &  Stability \\ 
\hline
CB 87	& 11.4 &	$(1.7\pm 0.2)\times 10^4$	& $2.73\pm 0.24$ & 9.6 &  stable \\ \hline
CB 110 & 21.8 & $(1.5\pm 0.6)\times 10^5$ & $7.21\pm 1.64$ & 8.5  &  unstable \\ \hline
CB 131 & 25.1 & $(2.5\pm 1.3)\times 10^5$ & $7.83\pm 2.35$ & 8.1  &  unstable \\ \hline
CB 134 & 13.2 & $(7.5\pm 3.3)\times 10^5$ & $1.91\pm 0.52$ & 1.8  &  unstable \\ \hline
CB 161 & 12.5 & $(7.0\pm 1.6)\times 10^4$ & $2.79\pm 0.72$ & 5.4   &  unstable \\ \hline
CB 184 & 15.5 & $(3.0\pm 0.4)\times 10^4$ & $4.70\pm 1.76$ & 11.4  &  unstable \\ \hline
CB 188 & 19.0 & $(1.2\pm 0.2)\times 10^5$ & $7.19\pm 2.28$ & 7.7 &  unstable \\ \hline
FeSt 1-457 & 10.9 & $(6.5\pm 1.7)\times 10^5$ & $1.12\pm 0.23$ & 1.4  &  unstable \\ \hline
Lynds 495 & 12.6 & $(4.8\pm 1.4)\times 10^4$ & $2.95\pm 0.77$ & 6.6  &  unstable \\ \hline
Lynds 498 & 11.0 & $(4.3\pm 0.5)\times 10^4$ & $1.42\pm 0.16$ & 5.7 & stable \\ \hline
Coalsack & 15 & $(5.4\pm 1.4)\times 10^4$ & $4.50$ & 8.1 & stable \\ \hline
\end{tabular}
\caption{Kinetic temperature, particle number density, mass, Jeans mass, and observed stability for several Bok globules \cite{Kan,Vain}.}
\label{tab:globules}
\end{table}

One may observe from Table \ref{tab:globules} that 7 out of the 11 considered Bok globules are predicted to be stable while observation reveals that they exhibit star formation. As the criterion (\ref{m21}) allows for critical masses smaller than the usual Jeans mass, it may potentially account for this discrepancy. For that, we impose that the critical mass of the model is equal to the mass of the Bok globule; this provides the maximal value of the critical mass in order to account for the data. We set $U_0 = \sigma^2$ (Virial theorem), and we assume that the ratio $\rho_0^d/ \rho_0^b$ corresponds to the ratio of the density parameter $\Omega_d / \Omega_b \approx 5.5$ today, as it has not changed that much during the evolution of the Universe (see e.g., \cite{b5,our}). With these assumptions, we are left with a single free parameter, i.e., $\sigma_d^2$. The lower values of $\sigma_d^2$ allowing to match with the observational data are given in Table \ref{tab2}, for each Bok globule.

\begin{table}[h!]
\scriptsize
\begin{tabular}{|l|c|c|c|c|c|c|} \hline
Bok Globule & $\sigma_b^2 [ 10^4 m^2 /s^2]$  & $n_{H_2} \text{[cm$^{-3}$]}$ & $M [M_{\odot}]$  & $M_J [M_{\odot}]$ & $\sigma_d^2 [ 10^6 m^2 /s^2]$  \\ 
\hline
CB 110 & $17.98$ & $(1.5\pm 0.6)\times 10^5$ & $7.21\pm 1.64$ & 8.5  & $8.53$  \\ \hline
CB 131 & $20.70$ & $(2.5\pm 1.3)\times 10^5$ & $7.83\pm 2.35$ & 8.1  & $49.81$  \\ \hline
CB 161 & $10.31$ & $(7.0\pm 1.6)\times 10^4$ & $2.79\pm 0.72$ & 5.4   & $1.02$ \\ \hline
CB 184 & $12.78$ & $(3.0\pm 0.4)\times 10^4$ & $4.70\pm 1.76$ & 11.4  & $0.87$  \\ \hline
CB 188 & $15.67$ & $(1.2\pm 0.2)\times 10^5$ & $7.19\pm 2.28$ & 7.7 & $18.44$  \\ \hline
FeSt 1-457 & $8.99$ & $(6.5\pm 1.7)\times 10^5$ & $1.12\pm 0.23$ & 1.4  & $3.08$  \\ \hline
Lynds 495 & $10.39$ & $(4.8\pm 1.4)\times 10^4$ & $2.95\pm 0.77$ & 6.6  & $0.80$  \\ \hline
\end{tabular}
\caption{$\sigma_b^2$, particle number density, mass, and Jeans mass for 7 of the Bok globules of Table \ref{tab:globules}, whose predicted stability is contradicted by observation, together with the saturation bounds for $\sigma_d^2$ obtained with Eq. (\ref{m21}).}
\label{tab2}
\end{table}

One may see that the observational stability is correctly accounted for, with $\sigma_d^2 \sim 10^6 m^2/s^2$. One may ask whether such values of $\sigma_d^2$ are physically well-motivated and whether they match with what is known about the properties of dark matter. To see that, one may use the observational evidence that there are no dark matter halos with a radius smaller that $R \sim 1kpc$ and a mass smaller than $M \sim 10^8 M_{\odot}$ \cite{Chavanis}. These ultracompact dark matter halos correspond typically to dwarf spheroidal galaxies like \textit{Fornax}. If we assume that Fornax is the smallest halo observed in the Universe and associate their mass and radius to the Jeans mass and Jeans length, we find $\sigma_d^2 \sim 0.13 \times 10^6 m^2/s^2$, which is comparable to the values of $\sigma_d^2$ given in Table \ref{tab2}, required to account for the stability of Bok globules.

\section{Conclusions}\lb{sec6}

In this work, we have analysed the Jeans-type gravitational instability for a mixture of baryonic and dark matter particles, in the post-Newtonian approximation. We have laid out a kinetic model consisting of two post-Newtonian collisionless Boltzmann equations, for each particle species, and the post-Newtonian Poisson equations. The relativistic Maxwell–Jüttner distribution function was used to evaluate the  components of the energy–momentum tensor in the Poisson equations. By considering perturbations, around background state, in the form of plane waves, we have established the post-Newtonian dispersion relation, for a mixture of baryonic and dark matter particles. This leads to a Jeans mass smaller than the standard Jeans mass, meaning that smaller masses are required to initiate the gravitational collapse. 

We have used this to study to what extent the model can account for the observational stability of Bok globules. In particular, a set of Bok globules are theoretically predicted as stable (using the standard Jeans mass), yet they are observed to exhibit star formation \cite{Kan}. We have shown that the present model can correctly account for this discrepancy, with physically reasonable values of the dark matter velocity dispersion. 
  
\section*{Acknowledgments}  (GMK) was supported by Conselho Nacional de Desenvolvimento Cient\'{i}fico e Tecnol\'{o}gico (CNPq), grant No.  304054/2019-4.

\end{document}